\documentclass[twocolumn,aps,floats,letterpaper,floatfix,groupedaddress]{revtex4-1}
\usepackage{graphicx}
\usepackage{dcolumn}
\usepackage{amsmath}
\usepackage{amsfonts}
\usepackage{amssymb}
\usepackage{bm}
\usepackage{epsfig,float,afterpage}
\usepackage{natbib}

\newcommand{\beq}{\begin{equation}}
\newcommand{\eeq}{\end{equation}}
\newcommand{\bes}{\begin{subequations}}
\newcommand{\ees}{\end{subequations}}
\newcommand{\bea}{\begin{eqnarray}}
\newcommand{\eea}{\end{eqnarray}}
\newcommand{\ba}{\begin{array}}
\newcommand{\ea}{\end{array}}
\newcommand{\beqn}{\begin{eqnarray*}}
\newcommand{\eeqn}{\end{eqnarray*}}

\newcommand{\f}[2]{\frac{#1}{#2}}

\newcommand{\om}{\omega}

\newcommand{\la}{\langle}
\newcommand{\ra}{\rangle}

\def\nn{\nonumber}

\newlength{\sizeonefig}
\newlength{\sizetwofig}
\begin{document}

\title{Probing Many-Body Localization by Spin Noise Spectroscopy}

\author{Dibyendu Roy, Rajeev Singh, Roderich Moessner}

\affiliation{Max Planck Institute for the Physics of Complex Systems,
  N{\"o}thnitzer Str. 38, 01187 Dresden, Germany}


\begin{abstract}
We propose to apply spin noise spectroscopy (SNS) to detect many-body
localization (MBL) in disordered spin systems. The SNS methods are relatively
non-invasive technique to probe spontaneous spin fluctuations. We here show that the spin noise signals obtained by cross-correlation SNS with two probe beams can be used to separate the MBL
phase from a noninteracting Anderson localized phase and a delocalized
(diffusive) phase in the studied models for which we numerically calculate
real time spin noise signals and their power spectra. For the archetypical case
of the disordered XXZ spin chain we also develop a simple phenomenological model.  

\end{abstract}
\vspace{0.0cm}
\maketitle
The fate of Anderson localization in the presence of inter-particle
interactions in a
disordered quantum medium is an exciting frontier in condensed matter physics \cite{Fleishman80, Altshuler97, Jacquod97, Gornyi05,
  Basko06, Oganesyan07, Znidaric08, Karahalios09, Pal10, Bardarson12, Vosk13,
  Serbyn13, Huse13, Imbrie14, Kjall14, Vasseur15, Ros15, Schreiber15, Bera15}. 
It is well known that all states of a one-dimensional (1D) disordered
chain of noninteracting particles are Anderson localized (AL) for any amount
of disorder \cite{Anderson58, Mott61}. Thus, the AL state is a perfect insulator as long as the particles are not coupled to
other degrees of freedom. In the presence of inter-particle
interactions, a dynamical transition from a delocalized (diffusive) phase to a
many-body localized (MBL) phase  has been predicted in disordered quantum media
when the strength of (quenched) randomness is increased \cite{Gornyi05, Basko06, Oganesyan07, Znidaric08,
  Karahalios09, Pal10}.  

The MBL state is also a perfect insulator, but it has different dynamical properties compared to the
AL state of noninteracting particles \cite{Bardarson12,
  Vosk13, Serbyn13, Vasseur15}. For example, entanglement entropy in the MBL
phase shows a slow logarithmic growth following a global quench in
an isolated system \cite{Bardarson12, Vosk13, Serbyn13}. However, the entanglement entropy is
very difficult to measure experimentally, and electrons or spins in
conventional solid-state systems are coupled to an environment such as a phonon bath.
  Recent studies show that signatures of the MBL phase can
  survive in the presence of weak coupling to a thermalizing environment
  \cite{Nandkishore14, Johri15}. In particular, the spectral functions of local
  operators can  be used to identify the MBL phase in the presence of weak dissipation \cite{Nandkishore14, Johri15}.

A new experimental approach based on a modified nonlocal spin-echo
protocol along with a double electron-electron resonance technique in
electron spin resonance has been proposed to distinguish the MBL phase from a
noninteracting AL phase and a delocalized phase at infinite
temperature \cite{Serbyn14}.  The proposed approach
can probe interaction effects, thus is able to separate the MBL phase from the
AL phase. The method does involve optical
pumping or polarization of local spins by external pulse fields, which can in
principle lead
to unwanted local heating and excitations. Here we propose a relatively
non-invasive method based on spin noise spectroscopy (SNS) to distinguish the
MBL phase from the AL phase and the delocalized phase in
disordered spin systems.

\begin{figure}
\includegraphics[width=\columnwidth]{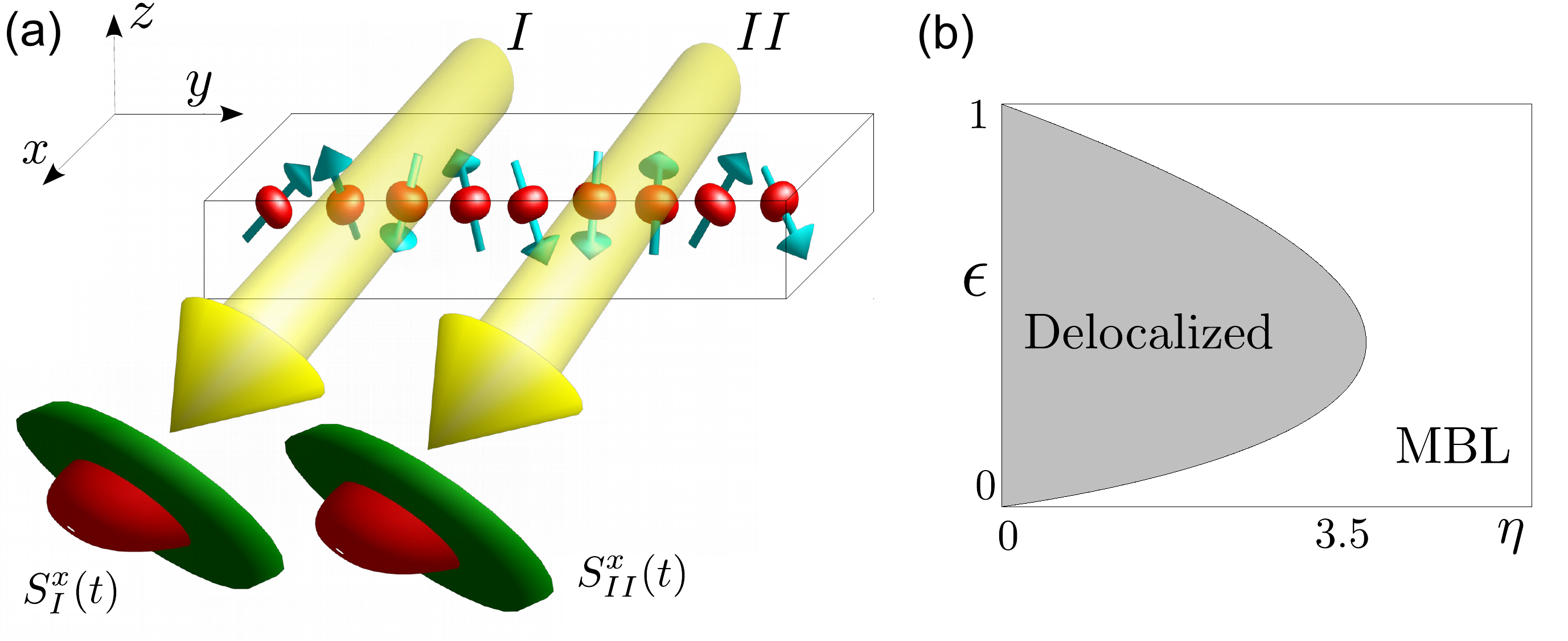}
\caption{(a) Schematic of the two-beam spin noise spectroscopy set-up with two
probe lasers detecting spin fluctuations of different spins. (b) Disorder
($\eta$) vs. energy density ($\epsilon$) phase-diagram of the disordered
Heisenberg chain \cite{Luitz15}.}
\label{schematic}
\end{figure}

The optical SNS method has been developed recently as an alternative to
conventional perturbation-based (pump-probe) techniques for measuring
dynamical spin properties \cite{Zapasskii13, Muller10}. Intrinsic spin fluctuations of electrons and holes  are passively detected in SNS  by measuring optical Faraday
rotation fluctuations of a linearly polarized probe laser beam passing through
the sample \cite{Aleksandrov81, Kuzmich99, Takahashi99, Crooker04,
  Oestreich05, Muller08, Crooker09,Crooker10, Li12, Berski13, Poltavtsev14,
  Glasenapp14}.  SNS with a
single probe laser has been useful in characterizing various properties (e.g., g-factors, relaxation rates and decoherence times) of
different spin ensembles, such as specific alkali atoms \cite{Crooker04},
itinerant electron spins in semiconductors \cite{Crooker09} and localized hole
spins in quantum dot ensembles \cite{Crooker10}. Last year, an extension of the traditional SNS has been proposed
and demonstrated by using two linearly polarized probe lasers for
detecting inter-species spin interactions in a heterogeneous two-component
spin ensemble interacting via binary exchange
coupling in thermal equilibrium \cite{Yang14, Roy15}. In this cross-correlation SNS method,  intrinsic spin
fluctuations from two different species are independently detected, and interaction effects are determined by the cross-correlation of these two spin
noise signals. Interaction effects between spins of a single species in a
sample can also be determined by two-beam SNS when the two probe laser
beams are spatially separated as shown in Fig.~\ref{schematic}(a)
\cite{Pershin13, Yang14}. Thus, we propose that the two-beam SNS measurements
can be used to separate the MBL phase from an AL
phase. In fact, as we show in our example, the responses from single-beam and two-beam
SNS can be efficiently employed to distinguish the MBL phase from both the
AL and delocalized phases.

In order to demonstrate our idea we study the spin-noise signals of a 1D
random-field XXZ spin chain, perhaps the best-studied ``canonical'' model
system for MBL. In addition, results for the disordered transverse-field Ising
model are given in the Suppl. Mat. \cite{Suppl15}.  The Hamiltonian of
disordered XXZ spin chain is given by
\bea
H=\sum_{i=1}^{N}J_{\perp}(S^x_iS^x_{i+1}+S^y_iS^y_{i+1})+J_zS^z_iS^z_{i+1}+h_iS^z_i,\label{Ham}
\eea
where the random fields $h_i$ are chosen from a uniform distribution within the
window $[-\eta,\eta]$, and $S^{x,y,z}_{N+1}=S^{x,y,z}_1$ for periodic boundary
conditions. 
 For isotropic exchange interactions ($J_z=J_{\perp}$), this is a disordered Heisenberg chain. The spin Hamiltonian in Eq.~\ref{Ham} can be mapped to a model of
interacting spinless Jordon-Wigner fermions on a lattice where $J_{\perp}$
represents hopping between neighboring sites, $J_z$ is inter-particle
interaction strength and $h_i$ denotes a random local chemical potential
\cite{Giamarchi04}. For $J_z=0$, the Hamiltonian in Eq.~\ref{Ham} represents noninteracting  fermions subject only to
a random local potential, and hence can be used to study the AL phase of
noninteracting particles. This model with a non-zero $J_z$ has been
extensively studied recently in the context of MBL. A disorder versus energy
density phase-diagram separating the delocalized and MBL phases of an isolated
chain is shown in Fig.~\ref{schematic}(b) following Ref.~\cite{Luitz15} where
 $\epsilon=(\la H \ra -E_{\rm max})/(E_{\rm min}-E_{\rm max})$ with $E_{\rm
   max}$ ($E_{\rm  min}$) being maximum (minimum) energy.   

For the Hamiltonian in Eq.~\ref{Ham}, we now describe the signals measured in the SNS set-ups. The measured response of the spin component tranverse to the
 random magnetic field in the single-beam SNS is 
\bea
\mathcal{C}^x_I(t)=\la\la \{S^x_I(t),S^x_I(0)\}\ra\ra, \label{sbSNS}
\eea
where $\{.,.\}$ is the anti-commutator and $\la\la ..\ra \ra$
denotes both (canonical) thermal and disorder averaging. Here, the $x$-component
of total spin polarization at the measurement spot at time $t$ is $S^x_I(t)=\sum_{l\in I}S^x_l(t)$ where the
sum runs over all local spin sites within the sample spot illuminated by the 
probe laser beam $I$. The correlation function $\mathcal{C}^x_I(t)$ describes the relaxation
of spontaneous spin fluctuations at the spot of the sample probed by the
single-beam SNS. The single-beam (local) spin-noise power spectrum is obtained
 by its Fourier transform,
\bea
P^x_{I}(\om)=\int_{-\infty}^{\infty}dt~ e^{i\om t} \mathcal{C}^x_I(t).
\eea  
Let $|E_n\ra$ denote a many-body eigenstate with eigenvalue $E_n$,
 $ H|E_n\ra=E_n|E_n\ra$. The single-beam spin-noise power spectrum reads,
\bea
P^x_{I}(\om) &=&\Big \la \f{1}{\mathcal{Z}}\sum_{n,p} \delta(\om+E_n-E_p) (e^{-\beta
  E_n}+e^{-\beta E_p})\nn\\ &&~~~~~~~~~~|\la E_n|S^x_I|E_p \ra|^2 \Big \ra,\label{sbpSNS}
\eea
with partition function $\mathcal{Z}=\sum_ne^{-\beta E_n}$,
and inverse temperature $\beta=1/k_BT$.  $\la..\ra$
in Eq.~\ref{sbpSNS} describes averaging over different disorder
realizations. The cross-correlation signal of the two-beam SNS is 
\bea
\mathcal{C}^x_{\rm cr}(t)=\la\la \{S^x_I(t),S^x_{II}(0)\}+\{S^x_{II}(t),S^x_{I}(0)\}\ra\ra,\label{tbSNS}
\eea
where $S^x_{II}(t)=\sum_{m \in   II}S^x_m(t)$, and the sum over $m$ runs
through all sites which the probe laser $II$ illuminates. The
cross-correlation spin noise power  
\bea
P^x_{\rm cr}(\om)&=&\int_{-\infty}^{\infty}dt~ e^{i\om t} \mathcal{C}^x_{\rm cr}(t)\\
&=&\Big \la \f{2}{\mathcal{Z}}\sum_{n,p} \delta(\om+E_n-E_p) (e^{-\beta
  E_n}+e^{-\beta E_p})\nn\\ &&~~~~~~~~~~{\rm Re}\big(\la E_n|S^x_I|E_p \ra  \la E_p|S^x_{II}|E_n \ra \big) \Big \ra.\label{tbpSNS}
\eea
   
We calculate the eigenvalues and eigenstates of the Hamiltonian using exact
diagonalization, and evaluate the spin noise signals in
Eqs.~\ref{sbSNS},\ref{tbSNS} and the corresponding power spectra in
Eqs.~\ref{sbpSNS},\ref{tbpSNS}. In the following we present results
using periodic boundary conditions at  high temperature,
$k_BT=50J_{\perp}$, averaging over 3000 disorder realizations. However, the
main results of this paper are also valid at moderate temperatures \cite{Suppl15}. We quote $J_z,h_i,\eta$ in the units of
$J_{\perp}$, and fix $J_{\perp}=1$ throughout the paper. 

\begin{figure}
\includegraphics[width=\columnwidth]{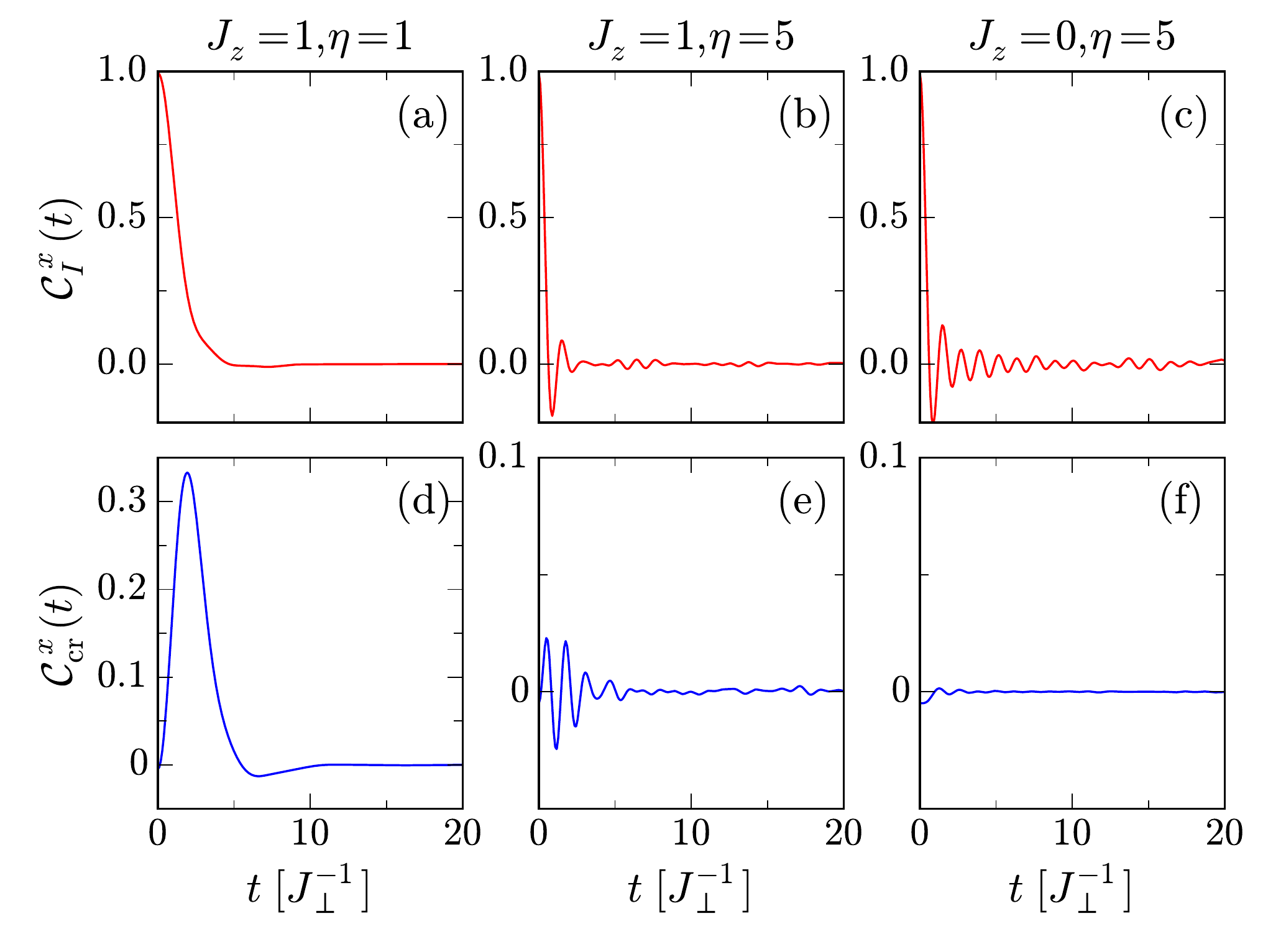}
\caption{Single-beam and two-beam spin noise responses in real time in the three different phases. (a-c)
  $\mathcal{C}^x_I(t)$  and (d-f) $\mathcal{C}^x_{\rm cr}(t)$  are obtained by exact
  diagonalization in the delocalized
  $(J_z=1, \eta=1)$, the MBL $(J_z=1, \eta=5)$ and the
  Anderson localized $(J_z=0, \eta=5)$ phase. Here $N=12$ and the beam $I$ and $II$ illuminate respectively spins 5,6
  and 7,8.}
\label{snstime}
\end{figure}

The single-beam spin noise response in real time for the interacting delocalized phase at low disorder
$(\eta=1)$ and for the MBL phase at high disorder $(\eta=5$, at which every
many-body eigenstate of the isolated chain is localized)  of the random
XXZ model is shown in Fig.~\ref{snstime}(a,b).  The
single-beam responses are clearly different in the two phases. The transverse spin component
relaxes exponentially in the delocalized phase, while in the MBL phase it appears to do so algebraically, and it also oscillates. We show the
 single-beam spin noise response of the AL phase $(J_z=0)$ in
 Fig.~\ref{snstime}(c), which again indicates an algebraic relaxation of the transverse spin component. The relaxation in this
 phase can be approximated as $\sin(t/\tau_A)/(t/\tau_A)$ with some
 characteristic  time scale $\tau_A$. We  use high disorder $(\eta=5)$ in Fig.~\ref{snstime}(c) to ensure that all states are
localized even for a finite-length noninteracting chain. We provide log scale
plots in the Suppl. Mat. \cite{Suppl15} for the real-time spin noise
responses in Fig.~\ref{snstime} to highlight the nature of their respective decays.

Within the single-beam SNS measurements, the slow spin relaxation due to
 interactions in the highly disordered XXZ model is masked by other mechanisms of relaxation (e.g., due to random fields and hopping) also
present in the noninteracting case. Thus, it becomes
difficult to separate the MBL phase from the AL phase by
 single-beam SNS. However,  interaction effects are nicely diagnosed by two-beam SNS when we cross correlate two different noise signals to exclude their
self-correlations. 

The two-beam cross-correlation spin noise responses in real time for the
three different phases are shown in Fig.~\ref{snstime}(d,e,f) when two beams
are just next to each other without overlapping (i.e., the separation between
the centers of the two beams is two lattice spacings). The responses here are
quite different in all three phases. A large cross-correlation between the
transverse components of different spins is developed in the diffusive phase,
and it relaxes exponentially fast. The cross-correlation in the MBL phase is relatively
smaller and it relaxes slowly with many oscillations. In the AL phase, the
transverse component hardly shows any cross-correlation between different
spins at high temperature. Notice that $\mathcal{C}^x_{I}(0)=1$  while
$\mathcal{C}^x_{\rm cr}(0)\approx 0$ in Fig.~\ref{snstime}, because $\la
S^x_l(0)S^x_m(0)\ra=\delta_{lm}/4$ at high temperature \cite{area}. We also perform these calculation for the
relaxation of spins along the $z$-direction, and find that it is 
similar in the MBL and AL phases even within the two-beam SNS measurements
\cite{Suppl15, BarLev15}. Thus, $z$-direction response is not a good candidate to separate these two phases.  

\begin{figure}
\includegraphics[width=\columnwidth]{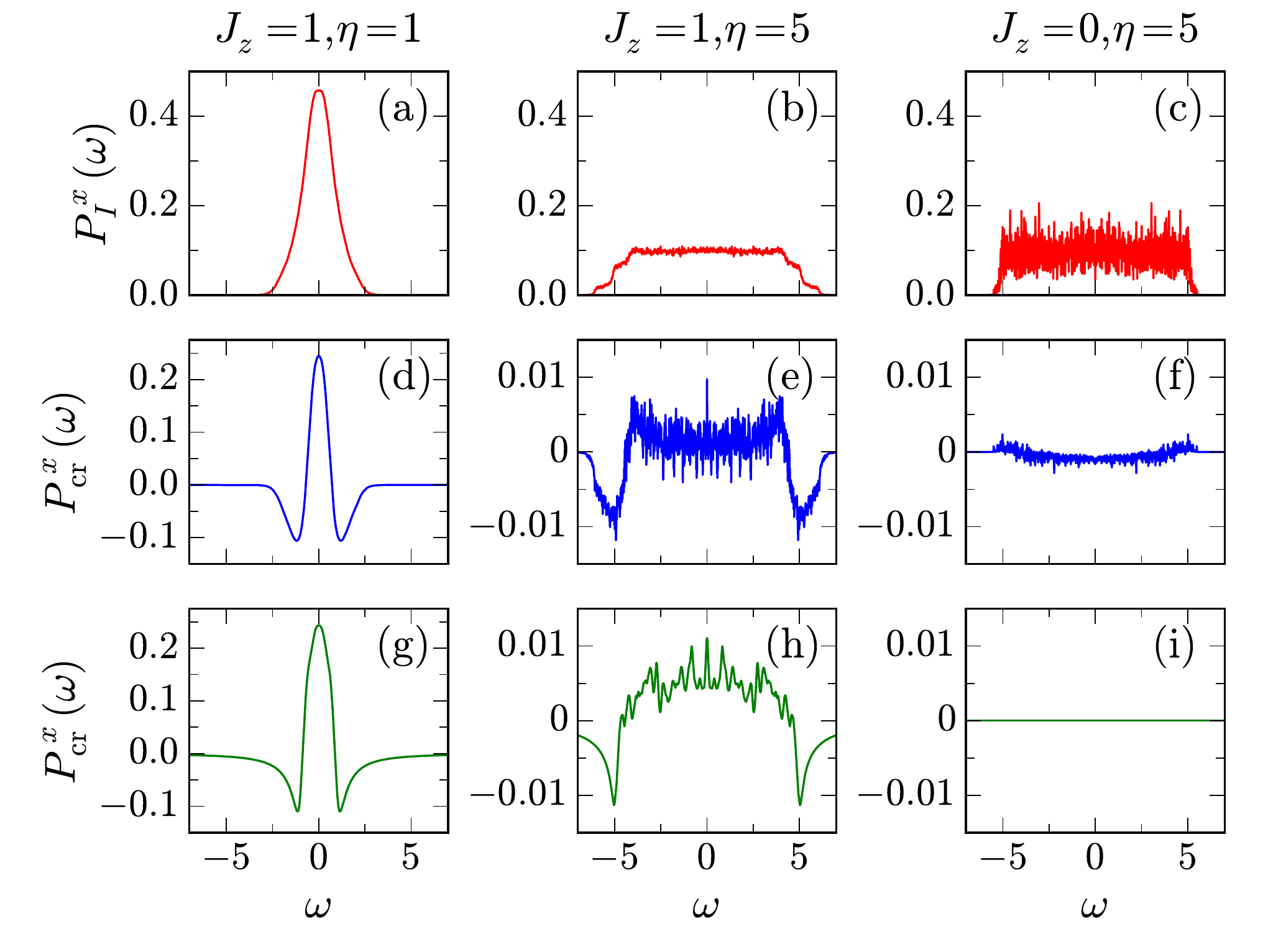}
\caption{Single-beam and two-beam spin noise power spectra in the three different
  phases. (a-c) $P^x_{I}(\om)$, (d-f) $P^x_{\rm cr}(\om)$ are obtained by exact
  diagonalization and (g-i) $P^x_{\rm cr}(\om)$ are calculated from Eq.~\ref{tbpSNS1} in the delocalized
  $(J_z=1, \eta=1)$, the MBL $(J_z=1, \eta=5)$ and the
  Anderson localized $(J_z=0, \eta=5)$ phase. We use
  $\gamma=0.068,\gamma_1=0.18$ in (g),  $\gamma=0.03,\gamma_1=0.08$ in (h) and
      $\gamma=0$ in (i). The parameters in (a-f) are the same as in Fig.~\ref{snstime}.}
\label{snsfreq}
\end{figure}

Turning to the frequency domain, the disorder averaged spin noise power spectra
of single-beam and two-beam SNS are shown in Fig.~\ref{snsfreq}. We find a
Lorentzian line-shape for the single-beam noise power spectrum $P^x_{I}(\om)$
[Fig.~\ref{snsfreq}(a)] in the diffusive regime reflecting the exponential spin relaxation. In the MBL phase, the shape of $P^x_{I}(\om)$ exhibits a plateau as shown in
Fig.~\ref{snsfreq}(b). The overall rectangular shape of
$P^x_{I}(\om)$ in the AL regime [Fig.~\ref{snsfreq}(c)] is somewhat
similar to that of the MBL phase but it shows much higher fluctuations. 

Crucially, the cross-correlation spin noise power spectra $P^{x}_{\rm cr}(\om)$
of the two-beam SNS in Fig.~\ref{snsfreq}(d-f)  differ significantly 
in shape and magnitude in the delocalized (diffusive), MBL and AL phases. The power spectrum in the delocalized phase 
[Fig.~\ref{snsfreq}(d)] can be perceived as the difference between two equal area Lorentzians of different widths. Thus, the
total area under the cross-correlation curve is zero, a signature of
high-temperature behaviour. The line-shape of $P^{x}_{\rm cr}(\om)$ in the MBL
phase is very different from that in the delocalized and AL phases. It is
also almost one order of magnitude smaller than in the delocalized phase but is clearly non-zero
[Fig.~\ref{snsfreq}(e)]. However, $P^{x}_{\rm cr}(\om)$ in the  AL phase is
 very small (one order of magnitude
smaller than that in the MBL phase at high temperature) and featureless
indicating essentially no correlation between spins. Note that $P^{x}_{\rm
  cr}(\om)$ shows both negative and positive values which signify (anti-)correlations between spins at different frequency/time scales. The anti-correlations in Fig.~\ref{snsfreq}(d,e) at higher frequencies
(shorter time) are induced by fast co-flips between different spins in the
presence of spin-exchange coupling. This feature is present in the
two-beam cross-correlation spectra in the delocalized and MBL phases but
is absent in the AL phase. 

 $P^{x}_{\rm cr}(\om)$ depends on the separation between
the two probe beams. With an increasing separation between the beams,
$P^{x}_{\rm cr}(\om)$ exhibits more oscillations in the delocalized and MBL
phases, and its magnitude falls rapidly in the MBL and AL phases as might be
expected in the presence of spatial localization. Also the shape of
$P^{x}_{\rm cr}(\om)$ in the three phases remains unchanged
when we explicitly couple the disordered XXZ chain to a weakly thermalizing environment. Both of these features are discussed in the Suppl. Mat. \cite{Suppl15}.

In order to understand  the cross-correlation noise
power spectra obtained from  numerics, we develop a simple
phenomenological model inspired by the experimental set-up of the two-beam
SNS. Let us denote the respective total spin
polarizations of the spins illuminated by the probe laser beams $I$
and $II$ by ${\bf S}_I$ and ${\bf S}_{II}$. We assume that the spins in beam $I$
 ($II$) can relax (a) due to spin-exchange interactions with the spins in beam
 $II$ ($I$), and (b) due to interactions with other spins in the
sample. The spin-exchange interactions between the spins from the two
illuminated spots conserve total spin but do transfer spins between the two
spots via the spin co-flips with a rate $\gamma$. This leads to
cross-correlations of their fluctuations.  For process (b) we define a net
spin relaxation rate $\gamma_{\alpha}$ for spins in beams $\alpha=I,II$. This
process does not conserve the total spin in the beams. Combining these with
Larmor precession in the random magnetic field, we obtain the following stochastic evolution equations \cite{Roy15}, 
\bea
\f{d{\bf S}_{\alpha}}{dt}&=&{\bf S}_{\alpha}\times {\bf h}_{\alpha}-\gamma_{\alpha}{\bf S}_{\alpha}-\gamma({\bf S}_{\alpha}-{\bf
  S}_{\bar{\alpha}})+{\boldsymbol \xi}_{\alpha},\label{SDP1}
\eea
where $\bar{I}=II,~\overline{II}=I$. Here we have included stochastic
fluctuations  ${\boldsymbol \xi}_{\alpha}$. For simplicity, we consider these
 as Gaussian white noise with  zero mean,
\bea
\la \xi_{\alpha i}(t) \xi_{\beta j}(t')\ra=\delta(t-t')\f{\delta_{ij}}{2}
\left( \delta_{\alpha \beta} \gamma_{\alpha} +\gamma( \delta_{\alpha \beta}-\delta_{\alpha \bar{\beta}}) \right),\label{FDrel}
\eea
with $i,j=x,y,z$.  The relation in Eq.~\ref{FDrel} ensures stationary
equal-time correlation $\la S_{\alpha   i}(t)S_{\beta j}(t)\ra \propto
\delta_{\alpha \beta}\delta_{ij}/4$  in the high temperature limit \cite{Roy15}. We coarse-grain the random magnetic fields within the spot of
each laser beams, and represent them by a single random field. Here we choose the random field along $z$-direction following
the Hamiltonian in Eq.~\ref{Ham}, i.e., ${\bf h}_{\alpha}=h_{\alpha}\hat{z}$ where $h_{\alpha}$ are two i.i.d. random
numbers chosen from a uniform distribution over $[-\eta,\eta]$. Averaging over noise, we find the cross-correlator of spin polarizations
in Eq.~\ref{tbpSNS} in compact form,
\bea
P^x_{\rm cr}(\om)=\gamma \sum \limits_{q=\pm} \Big\la \f{\chi_{q} }{\chi_{q}^2+\kappa_{q}^2}\Big\ra,   
\label{tbpSNS1}
\eea
where $\chi_{\pm}=\gamma_1\gamma_2-(\om \pm h_I)(\om \pm
h_{II}),~\kappa_{\pm}=\gamma_3(2\om\pm (h_I+h_{II}))$. Here
$\gamma_I=\gamma_{II}=\gamma_1$, $\gamma_2=\gamma_1+2\gamma$ and
$\gamma_3=\gamma_1+\gamma$ are characteristic rates that influence broadening
of the peaks.  $\la ..\ra$ in Eq.~\ref{tbpSNS1} denotes
averaging over the random fields $h_{\alpha}$. Due to the presence of the random field throughout the sample,
the phenomenological relaxation rates $\gamma, \gamma_I$ and $\gamma_{II}$ vary with disorder strength $\eta$, and we use them as fitting parameters. We plot the cross-correlator $P^x_{\rm cr}(\om)$ after
averaging over the random fields for $\eta=1,5$ in Fig.~\ref{snsfreq}(g,h). We
find remarkably good
agreement between the numerical results obtained at high temperature and
the results from our phenomenological model. From
Eq.~\ref{tbpSNS1}, $P^x_{\rm cr}(\om)=0$ for $\gamma=0$, in the AL phase
($J_z=0$).  %We note, however, that in this model we assume a direct and
            %instantaneous co-flip of spins between two illuminated spots
            %which can be justified only when two beams are spatially next to
            %each other without overlapping. When the illuminated spots are
            %spatially separated, the co-flip process  propagates through the
            %middle spins and it introduces a new time scale which is not
            %included in our simplified model. 
We note, however, that our model assumes instantaneous transfer of
polarization density through direct co-flips of spins, appropriate for
directly adjacent spots. If this condition is not fulfilled, a
delay for the propagation of polarization density over the distance between
the illuminated spots will have to be added to our simplified model. Secondly, our model works only at
relatively high temperatures and the results obtained from the
phenomenological model would deviate from the numerical results at low temperature \cite{Suppl15, Roy13}. 
 
To summarize, we have calculated spin noise signals in the diffusive, AL and
MBL  regimes of the  disordered XXZ and
transverse-field Ising spin chains, for different temperatures as well as beam separations. 
We have shown how these signals can be
used to identify the presence of MBL. 
The current proposal addresses
disordered spin systems beyond previously addressed regimes
of infinite temperature and strictly isolated systems. For an actual experiment, it
is necessary to have a disordered spin system whose
interaction energies and relaxation time scales are within the
bandwidth of the recent optical SNS measurements (nearly 100 GHz)
\cite{Berski13}. In this regard, solid state systems, such as
localized spin defects in solids (e.g., nitrogen-vacancy centers in
diamond) \cite{Serbyn14, Yao14, Childress06, Neumann10}, are suitable
candidates for SNS measurements to detect  MBL. Apart from SNS, spatially
resolved spin noise from quasi-1D or 2D layers of interacting spins can also
be measured using recently demonstrated sensitive nanoscale magnetic resonance
imaging using nitrogen-vacancy centers in diamond \cite{Myers14, Rosskopf14}.

%{\bf Acknowledgments}

We thank Vadim Oganesyan  and Shivaji Sondhi for helpful discussions.

\pagebreak
\onecolumngrid
\begin{center}
{\large \bf Supplemental Material for Probing Many-Body Localization by Spin
  Noise Spectroscopy} \\
\vspace{2mm}
{Dibyendu Roy, Rajeev Singh, Roderich Moessner}
\end{center}
\twocolumngrid
\setcounter{figure}{0}
\renewcommand\thefigure{S\arabic{figure}}
\setcounter{equation}{0}
\renewcommand\theequation{S\arabic{equation}}

\begin{figure}
\includegraphics[width=\columnwidth]{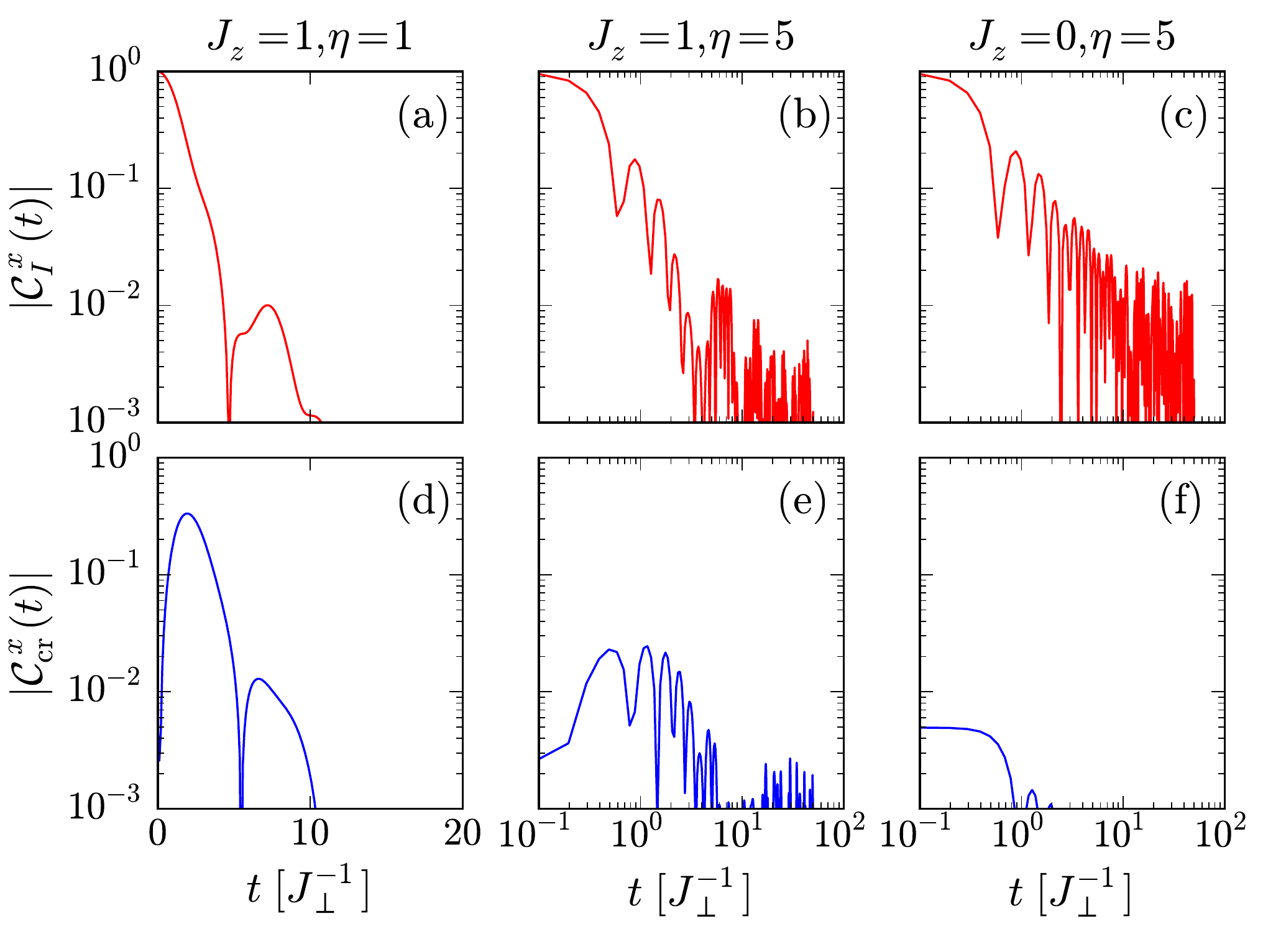}
\caption{Log scale plots of single-beam and two-beam spin noise
  responses of transverse spin component in real time in three different
  phases of disordered XXZ spin chain. (a,d) in log-linear scale in the delocalized phase
  $(J_z=1, \eta=1)$ and (b,c,e,f) in log-log scale in the MBL $(J_z=1,
  \eta=5)$ and the Anderson localized $(J_z=0, \eta=5)$ phase. Here $N=12$ and the beam $I$ and $II$ illuminate respectively spins 5,6
  and 7,8.}
\label{snstimexlog}
\end{figure}

We show plots in log-linear and log-log scales of real-time
spin noise responses of the transverse spin component of disordered XXZ
spin chain to elucidate the nature of their relaxation. We plot absolute values of
$\mathcal{C}^x_I(t)$ and $\mathcal{C}^x_{\rm cr}(t)$ in log scale and time $t$
in linear scale in Fig.~\ref{snstimexlog}(a,d). Both the real-time
responses in the delocalized (diffusive) phase appear to exhibit 
exponential decay in time. In Fig.~\ref{snstimexlog}(b,c,e,f) we plot both
 absolute values of $\mathcal{C}^x_I(t)$ and $\mathcal{C}^x_{\rm cr}(t)$
and time $t$ in log scale. While the 
single-beam spin noise response of the transverse spin component in real time shows algebraic (power-law)
decay both in the MBL and AL
phases, cross-correlation spin noise response shows nearly algebraic decay
only in the MBL phase.  The transverse spin component shows significantly less cross-correlation in the AL 
phase, as is evident in Fig.~\ref{snstimexlog}(f).

\begin{figure}[H]
%\vspace{5mm}
\includegraphics[width=\columnwidth]{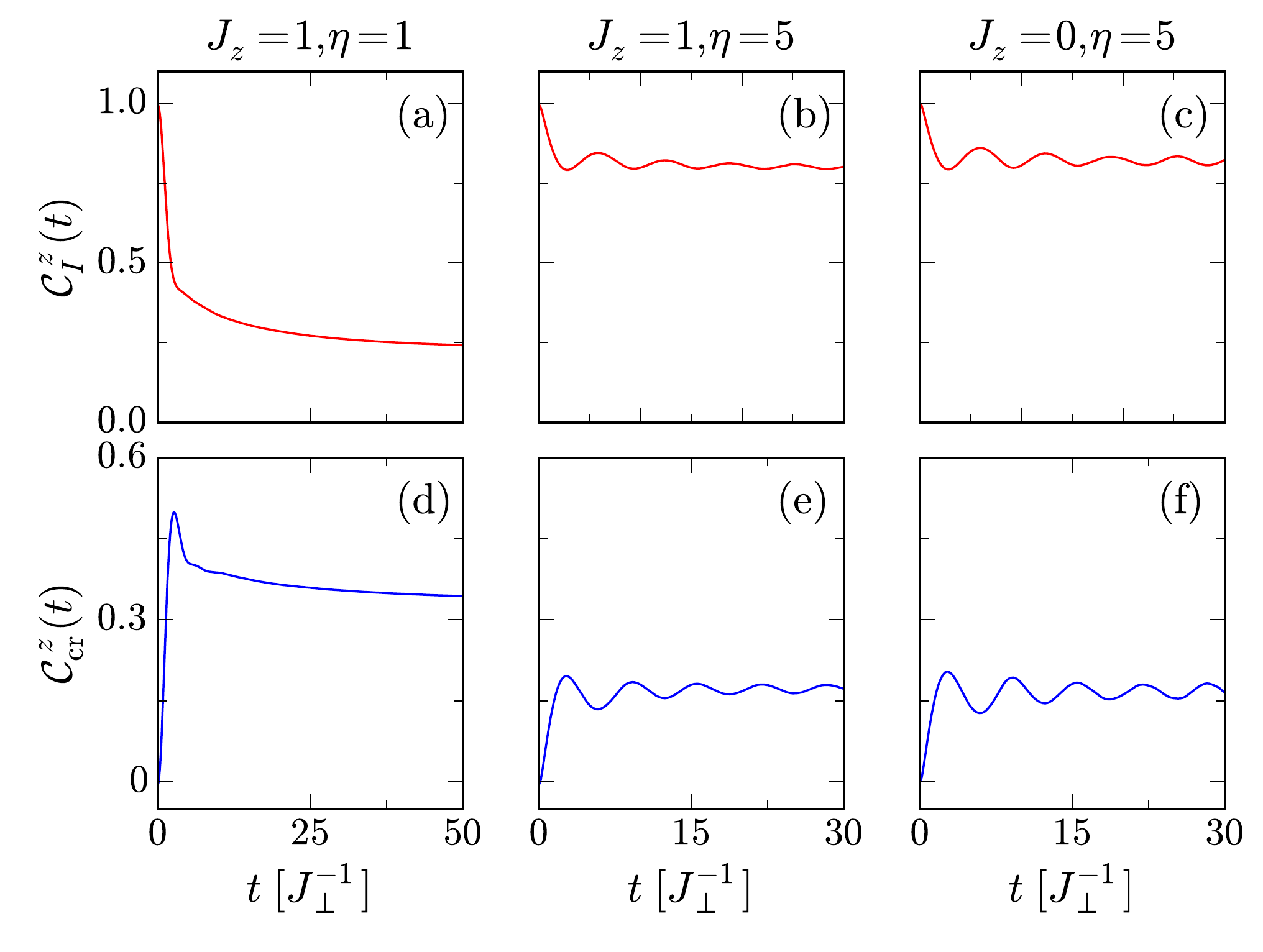}
\caption{Single-beam and two-beam spin noise responses of longitudinal spin
  component in real time in the delocalized, MBL and Anderson localized phases
  of disordered XXZ spin chain. The parameters are the same as in Fig.~\ref{snstimexlog}.}
\label{snstimez}
\end{figure}

In Fig.~\ref{snstimez} we show single-beam and two-beam spin noise
responses of the longitudinal spin component ($z$-direction) of disordered XXZ
spin chain in real time in the three
different phases. Both spin noise
responses show different relaxation behaviour
in the delocalized and the localized phases. However, the relaxation of the 
longitudinal spin component is quite similar in the MBL and the AL phases within
both single-beam and two-beam cross-correlation measurements. 

\begin{figure}[H]
\includegraphics[width=\columnwidth]{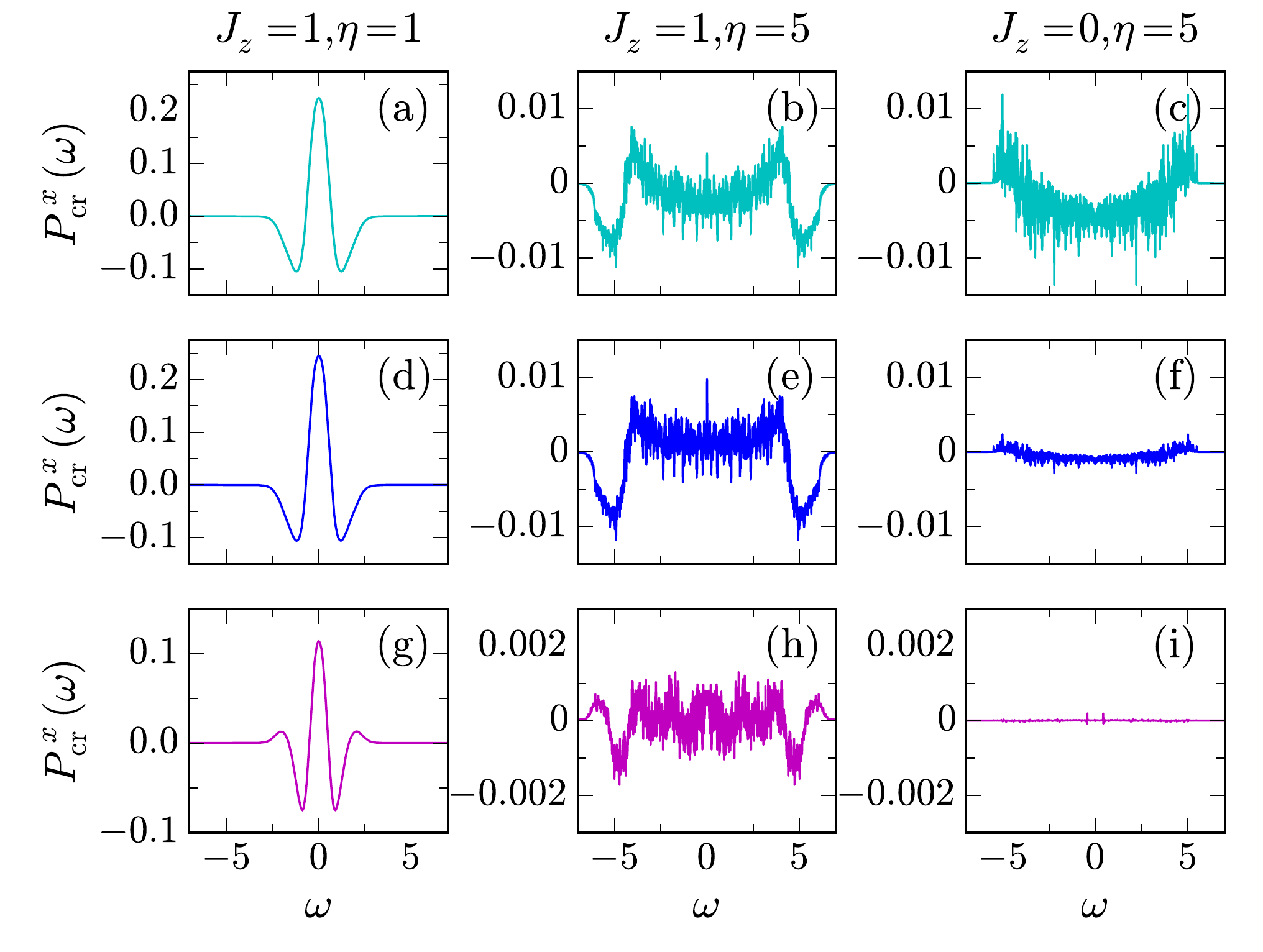}
\caption{Dependence of cross-correlation spin noise power spectra $P^x_{\rm
    cr}(\om)$ on temperature and separation between the probe beams in the
  delocalized, MBL and Anderson localized phases of the disordered XXZ spin
  chain. Here $N=12$ and temperature is $10J_{\perp}$ in (a-c) and
  $50J_{\perp}$ in (d-i). The beam $I$ and
  $II$ illuminate respectively spins 5,6 and 7,8 in (a-f) and spins 5,6
  and 8,9 in (g-i).}
\label{snsfreqt}
\end{figure}

Next, we discuss how cross-correlation spin noise spectra
$P^x_{\text cr}(\om)$ of transverse spin component of the disordered XXZ spin
chain depend on temperature and
  separation between the probe beams in the three different phases. For this we change temperature and
  separation between the probe beams from those in Fig.~3 of the main
  text. For comparison in Fig.~\ref{snsfreqt}(d-f) we use the same temperature and separation between the probe beams as in the main text.   
We reduce temperature to $10J_{\perp}$ in
  Fig.~\ref{snsfreqt}(a-c), and we find that the line-shapes of $P^x_{\text
    cr}(\om)$ in all three phases remains similar to that in Fig.~\ref{snsfreqt}(d-f) at high
  temperature $50J_{\perp}$. Thus, the two-beam SNS is still a
  good probe to separate the three different phases. However, fluctuations in $P^x_{\text
    cr}(\om)$ in the AL phase at low temperature are increased by one order of
  magnitude from  that at high temperature. In fact, $P^x_{\text cr}(\om)$ is
  similar in magnitude both in the MBL and AL phases but still has different
  shapes.  
In Fig.~\ref{snsfreqt}(g-i) we move the beam $II$ by one lattice spacing, and we  
 find extra oscillations in $P^x_{\text cr}(\om)$ compared to 
 Fig.~\ref{snsfreqt}(d,e) in the delocalized and MBL phases. We also notice that the magnitude of $P^x_{\text cr}(\om)$ is
 reduced by one order of magnitude in the MBL phase and by two orders in the AL phase
 compared to Fig.~\ref{snsfreqt}(e-f). We use periodic boundary
 conditions and averaging over 3000 disorder realizations in Figs.~\ref{snstimexlog},\ref{snstimez},\ref{snsfreqt}, except  in Fig.~\ref{snsfreqt}(h) where 10000 disorder realizations are used.

\begin{figure}
\includegraphics[width=\columnwidth]{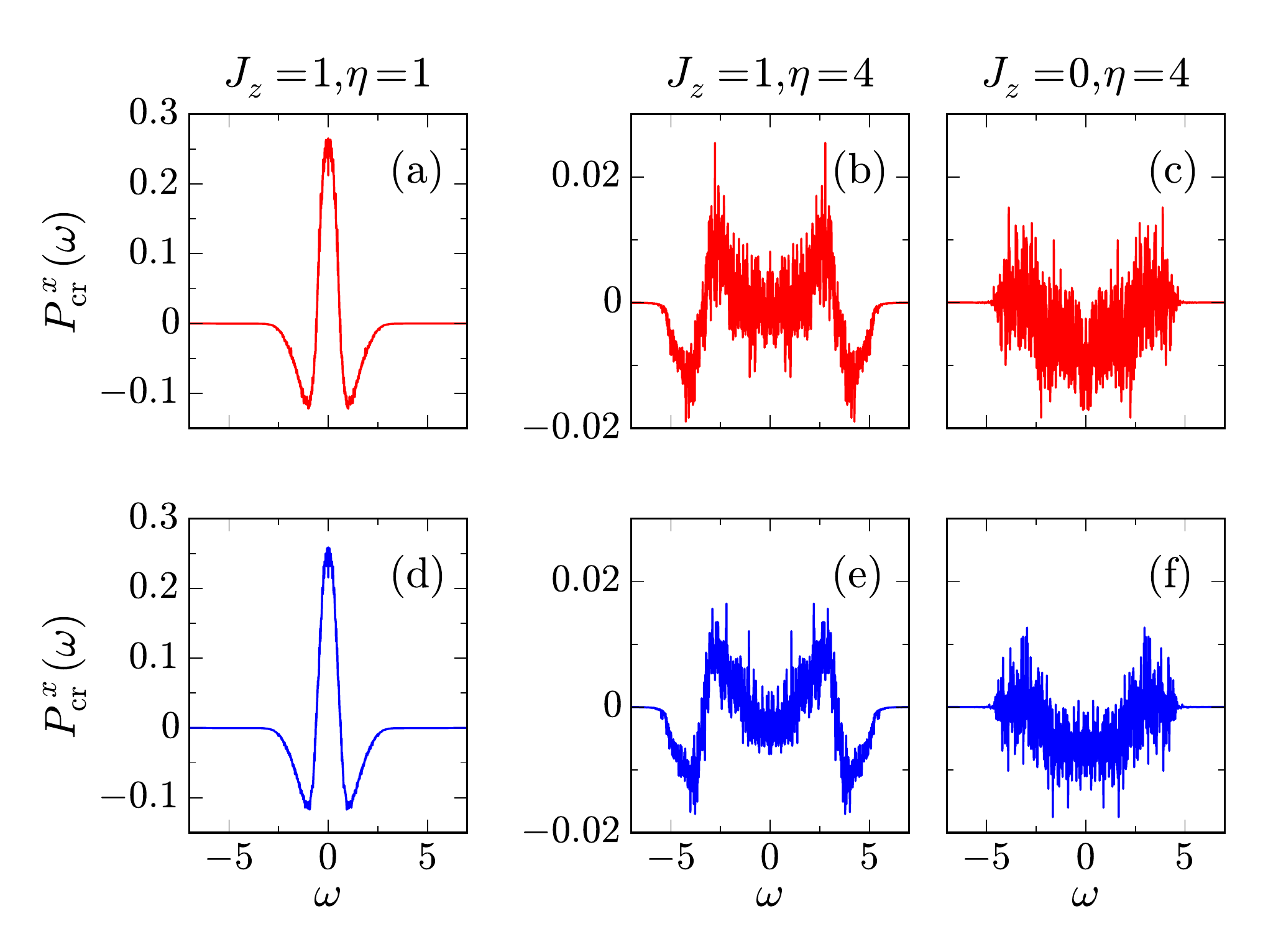}
\caption{Cross-correlation spin noise power spectra $P^x_{\rm cr}(\om)$ in the
  delocalized $(J_z=1, \eta=1)$, the MBL $(J_z=1, \eta=4)$ and the Anderson
  localized phases $(J_z=0, \eta=4)$ of the disordered XXZ spin
  chain coupled to a thermalizing bath. The bath coupling $J_c=0.15$ in
  (a-c) and $J_c=0.25$ in (d-f). Here $N=6$, $J_{\perp}=J_b=\eta_b=1$, 
  temperature is $50J_{\perp}$ and
  beams $I$ and $II$ illuminate respectively spins 2,3 and 4,5.}
\label{snsBath}
\end{figure}

We check the nature of cross-correlation spin noise power spectra of
disordered XXZ spin chain by explicitly coupling the chain (system) to a
thermalizing  bath. We model the bath here by  a weakly disordered XXZ spin
chain in the diffusive phase (nonintegrable). The Hamiltonians of the
bath and system-bath coupling are respectively,
\bea
H_{\rm
  bath}&=&\sum_{i=1}^{N}J_{b}\vec{\sigma}_i.\vec{\sigma}_{i+1}+h^b_i\sigma^z_i,\\
H_{\rm
  c}&=&\sum_{i=1}^{N}J_{c}\vec{S}_i.\vec{\sigma}_{i},
\eea
where the random fields $h^b_i$ are chosen from a uniform distribution within
the window $[-\eta_b,\eta_b]$, and $\vec{\sigma}_{N+1}=\vec{\sigma}_1$ for
periodic boundary conditions. Here $J_c$ determines strength of the system-bath
coupling, and the bath should be effective in thermalizing the system when the
coupling matrix element is of the order of the many-body level spacing in the
bath. The full Hamiltonian of the system plus bath is $H_{\rm f}=H+H_{\rm
  bath}+H_{\rm c}$  where $H$ is given in Eq.~\ref{Ham} of the main text. We
calculate spin noise signals in an eigenstate $|n\rangle$ of $H_f$ at an energy $E_n$
corresponding to that of the thermal ensemble at inverse temperature $\beta$,
$E_n=\la n|H_{\rm f}|n\ra={\rm Tr}[e^{-\beta H_{\rm f}}H_{\rm
  f}]/{\rm Tr}e^{-\beta H_{\rm f}}$. In
Fig.~\ref{snsBath} we show two-beam cross-correlation spin noise spectra of
the transverse spin component  in the three different phases for two different
values of $J_c$. Noise spectra are shown after averaging over $30000$
(except $10000$ in Fig.~\ref{snsBath}(a,d)) disorder realizations. We find that the shapes of two-beam
spin noise spectra in Fig.~\ref{snsBath} are similar to those obtained earlier
without explicitly coupling the system to a thermalizing bath. This shows
the robustness of two-beam spin noise signals to distinguish the three different
phases. We note from Fig.~\ref{snsBath} that the fluctuation in the noise
spectra falls with increasing bath coupling as long as the coupling is smaller
than the characteristic energy scales in the system. It signals better
thermalization with increasing bath coupling without destroying the signatures
of the three different phases in the isolated disordered XXZ chain. We mention
that the large fluctuations in the noise spectra shown in Fig.~\ref{snsBath} 
are also due to relatively shorter chain length which we were able to simulate for this case.

\begin{figure}
\includegraphics[width=\columnwidth]{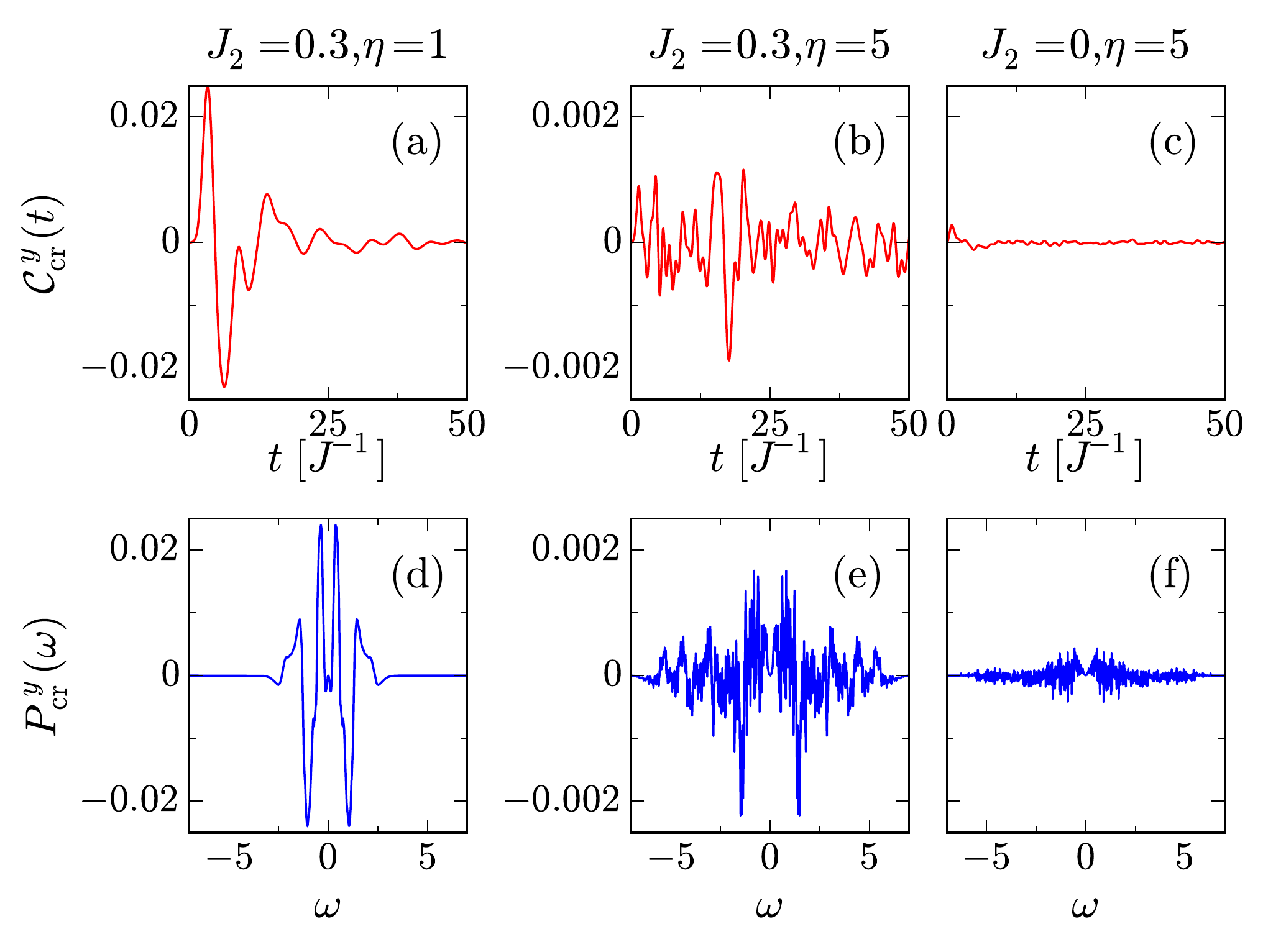}
\caption{Two-beam spin noise responses of transverse spin
  component in the three different phases of disordered transverse-field
  Ising chain. (a-c)
  $\mathcal{C}^y_{\rm cr}(t)$  and (d-f) $\mathcal{P}^y_{\rm cr}(\om)$  are obtained
  numerically in the delocalized   $(J_2=0.3, \eta=1)$, the MBL $(J_2=0.3,
  \eta=5)$ and the Anderson localized $(J_2=0, \eta=5)$ phase. Here $N=12$,
  $J=1$, $h=0.6$,   temperature is $50J$ and
  beams $I$ and $II$ illuminate respectively spins 5,6 and 7,8.}
\label{snsTI}
\end{figure}

Finally, we present cross-correlation spin noise signals measured by
two-beam SNS for a disordered
transverse-field Ising model with next-nearest neighbor
coupling. The Hamiltonian of this model is \cite{Kjalls14}
\bea
H_{\rm I}=-\sum_{i=1}^{N-1}J_i S_i^z S_{i+1}^z+J_2\sum_{i=1}^{N-2} S_i^z
S_{i+2}^z+h \sum_{i=1}^NS_i^x,
\eea
where the nearest neighbor couplings $J_i=J+\delta J_i$, with all random
$\delta J_i$ chosen from a uniform distribution within the window $[-\eta,
\eta]$; $h$ is a constant magnetic field. For finite $J_2$, the above
model has a delocalized phase at low disorder and an MBL phase at high
disorder \cite{Kjalls14}. The system is in the AL phase when $J_2=0$. We
numerically calculate
cross-correlation spin noise signals of the spin component along $y$-direction. We quote $J_2,h,\eta$ in the units of $J$, and
fix $J=1$. The cross-correlation spin noise signals in real
time in the delocalized, MBL and AL phases are shown in Fig.~\ref{snsTI}(a-c) at
high temperature. A large cross-correlation between the transverse spin
components of separate spins is developed in the delocalized phase and it
relaxes relatively fast, while the  cross-correlation is much weaker in the MBL
phase and it relaxes very slowly. The transverse spin component hardly shows any cross-correlation
between different spins at high temperature in the AL phase. We
plot cross-correlation spin noise power spectra in Fig.~\ref{snsTI}(d-f) which
are different in shape and magnitude in the three phases. Especially, the
cross-correlation power spectra in the MBL and AL phases are very
different in shape.  We use periodic boundary
 conditions, averaging over 3000 (in Fig.~\ref{snsTI}(a-c)), 6000  (in
 Fig.~\ref{snsTI}(d)) and 20000 (in Fig.~\ref{snsTI}(e,f)) disorder realizations.

\begin{thebibliography}{34}
\bibitem{area} In Figs.~\ref{snstime},\ref{snsfreq} we assume that two spins
  are illuminated by the each probe beam  which gives $\mathcal{C}^x_{I}(0)=\la \{S^x_5(0)+S^x_6(0),S^x_5(0)+S^x_6(0)\}\ra=1$.
\end{thebibliography}
\end{document}